\documentstyle[12pt,aps,epsf]{revtex}
\newcommand{\braket}[1]{ \langle #1 \rangle}
\newcommand{\bra}{ \langle }
\newcommand{\ket}{\rangle}

\newcommand{\bfl}{\begin{flushleft}}
\newcommand{\efl}{\end{flushleft}}
\newcommand{\bct}{\begin{center}}
\newcommand{\ect}{\end{center}}
\newcommand{\bfr}{\begin{flushright}}
\newcommand{\efr}{\end{flushright}}

\newcommand{\bv}[1]{\mbox{\boldmath $#1$}}
\newcommand{\dfrac}[2]{\frac {\strut #1}{\displaystyle{#2}}}
\def\NP#1#2#3{Nucl. Phys. {\bf #1} (#2) #3}
\def\PR#1#2#3{Phys.~Rev.~{\bf #1} (#2) #3}
\def\PRL#1#2#3{Phys.~Rev.~Lett. {\bf #1} (#2) #3}

\def\PTP#1#2#3{Prog. Theor. Phys. {\bf #1} (#2) #3}
\def\av#1#2#3{{\bf #1} (#2) #3}
\begin{document}
\draft
\tighten
\title{
  Monte-Carlo Study of Bound States in a Few-Nucleon System\\
  -- Method of Continued Fractions --}

\author{Hiroshi Masui\footnotemark[1]
  and Toru  Sato}
\address{Department of Physics, Graduate School of Science\\
Osaka University, Toyonaka, 560-0043, Japan}
\footnotetext[1]{\footnotesize
Present address:
{\it Department of Physics, Graduate School of Science
Hokkaido University, Sapporo, 060-0810, Japan}}
\date{\today}
\maketitle
\begin{abstract}
The aim of the present paper is to propose a new type of Monte-Carlo method 
which enables us to study the excited state
of many-body systems. 
It consists of the method of continued fractions (MCF) 
and a Monte-Carlo random walk for multi-dimensional integration. 
The convergence of the MCF and its accuracy are studied in detail 
for a one-dimensional model. 
As an application of this method to a realistic problem, 
we study the ground state 
of $^4$He by using the Volkov potential. 
It is shown that our results 
are in good agreement with those of previous works
obtained through different types of formalism. 
This implies that our new method  
works effectively in the study of many-body nuclear systems.
\end{abstract}
\section{Introduction}
The exact treatment of  nuclear many-body
systems is an  important benchmark 
for investigating 
the dynamics of multi-nucleon systems.
It is a testing ground of potential models and equations of motion,
such as those involving multi-nucleon interaction,
relativistic dynamics,
and possible modification of nucleon properties inside nuclei.

At present, various methods to solve many-body problems have been developed.
One of the standard methods is the Faddeev or
Faddeev-Yakubovsky (FY) equation.
The FY equation has been applied
to the investigation of the ground state of $^4$He
using a realistic nuclear potential.\cite{FY,FY2}
A common problem in  attempts to extend the `exact' study of many-body
systems is the rapid increase of the number of channels and dimensions
as the number of particles increases.
One way to overcome this problem is
to diagonalize the Hamiltonian
in a subspace spanned by 
appropriate  bases chosen using a stochastic method.
The other way is the direct calculation of the matrix elements
using a Monte-Carlo method.
In the Monte-Carlo method,
one can perform an effective multi-dimensional integration
by taking the importance sampling 
from the essential part of the phase space.

Extensive works have been carried out
by using a variational Monte-Carlo method (VMC)\cite{VMC} and
a Green's function Monte-Carlo method (GFMC).\cite{cal1,cal4}
VMC and GFMC have been applied to the study of 
the ground state properties of nuclei with mass number $A\leq7$
using realistic nuclear potential.\cite{cal4}
The bound state energies and also the electro-magnetic properties
of few-nucleon systems have been studied extensively.
In VMC, we can obtain the upper bound of the ground state energy
for which the matrix elements are evaluated by Monte-Carlo integration.
Simultaneously, we obtain an analytic form of the trial function
which describes the ground state quite well.
In GFMC, we start the calculation with a trial function
obtained by VMC, and the imaginary-time evolution  $e^{-\tau H}$
extracts the improved ground state wave function from the trial function. 
Here, the imaginary-time evolution is carried out by using
the quantum Monte-Carlo random walk method.
Due to the $e^{-\tau H}$ projection,
we only obtain the properties for
the lowest state of the system with a given quantum number.
Excited states in the diffusion Monte-Carlo method can be investigated
using a diagonalization procedure 
in the large space shell-model calculation with an auxiliary-field 
Monte-Carlo approach\cite{qmcd,kume}
and  in one- and two-dimensional examples of GFMC.\cite{cheon}
For the purpose of conducting  systematic studies in few-nucleon systems,
it is necessary to investigate excited states as well as
the ground state.
Therefore we should develop a method for this purpose
in the context of few-nucleon physics.

In this paper, we propose a new Monte-Carlo method to study
the bound state problems in few-nucleon systems.
Here, our purpose is
to find an effective method to extract eigenfunctions 
without referring to the imaginary-time evolution method,
but still using the Monte-Carlo random walk.
Therefore, the method is not restricted
to the ground state problem.
We use the method of continued fractions (MCF)
as a central formalism in our new calculation method.
MCF was originally proposed
by J. Hor\' a\v cek and T. Sasakawa \cite{mcf}
as a fast calculational method to evaluate
T-matrix elements and was applied to the three-nucleon system.
The wave functions which appear in the MCF iteration are
generated so as to be orthogonal to the lower order ones.
Then the matrix elements calculated by those wave functions rapidly converge.
Moreover, this method is simple and allows us to use any function
as a starting trial wave function and
to adopt a nuclear potential including a many-body potential.

The direct application of MCF with Monte-Carlo integration,
however, causes a few serious problems:
First, the original form of the MCF iteration formula \cite{mcf_bind}
is not suitable to apply a Monte-Carlo random walk,
since the modified Green's functions used
in each iteration are written in a separable form.
We find with a slight modification
that the iteration formula can be written
in a form similar to the diffusion equation.
There is, however, a crucial difference between them.
In our new iteration formula,
a higher order MCF wave function is generated
from a lower order one.
Therefore, the equation cannot be a simple iteration of
the same function as in the diffusion equation. 
Second, in order to guarantee the orthogonalization 
of each MCF wave function,
we must deal the difference of two MCF wave functions.
This causes a sort of the ``sign problem''.\cite{kalos2}
Here, we propose a correct method of counting the weight 
in the random walk procedure
for an equation of a non-iterative type.
As a result, we find out that our method is
closely related to ``the multiple cancellation'' method.\cite{cancel}
With our cancellation method, we can take care of the
sign problem in our formula.

The validity of our method of counting weight and
the convergence of the MCF are extensively examined
for a simple bound state problem in one-dimension
including the excited states.
As a more realistic example of a multi-dimensional problem, 
we study $A=4$ nuclei with the Volkov potential.\cite{volkov}
Our method is successful in the treatment both examples.

We briefly review the method of continued fractions in \S 2.
Our random walk method in a non-iterative equation and 
multiple cancellation mechanism are discussed in \S 3.
We present our results in \S 4, and a summary is given in \S 5.

\section{Method of continued fractions}
The method of continued fractions (MCF)
has been developed as an efficient method
to study scattering problems in atomic and
nuclear physics.\cite{mcf,mcf_bind}
We briefly review this method, here.

The Schr\"odinger equation of nuclear bound state can be
written as 
\begin{equation}
  | \phi \ket = G_0(E) V | \phi \ket
  \mbox{ ,}
  \label{schr_ls}
\end{equation}
where $G_0(E)$ is a free Green's function,
\begin{equation}
  G_0(E)=\frac{1}{E-H_0}
  \mbox{ ,}
\end{equation}
and $H_0$ and $V$ are the kinetic energy
of the nucleons and the nuclear potential, respectively.

In this paper,
we use the modified Green's function (MCFG) approach,\cite{mcf}
since it is easy to implement in Monte-Carlo evaluations.
We introduce the arbitrary functions $|F_0 \ket$ and $|f\ket$
which are not orthogonal
to the eigenfunction $| \phi \ket$.
A new function $|F_1 \ket$ is defined as  
\begin{equation}
  | F_1 \ket \equiv  G_0(E) V | F_0 \ket
  \mbox{ ,}
  \label{f_1} 
\end{equation}
and a modified Green's function $G_1(E)$ as
\begin{equation}
  G_1(E)  \equiv  G_0(E) -
  \frac{| F_1 \ket \bra f | }{\braket{f | V | F_0}}
  \mbox{ .}
  \label{g_1}
\end{equation}
$G_1(E)$ is chosen so as to operate only on the space
which is orthogonal to $V|F_0 \ket$ as
\begin{equation}
  G_1(E) V | F_0 \ket  =  0
\mbox{ .}
\end{equation}
We eliminate $G_0(E)$ from the Schr\"odinger equation (\ref{schr_ls})
using (\ref{g_1}):
\begin{equation}
  | \phi \ket  =  \left( G_1(E) +
    \frac{| F_1 \ket \bra f |}{\braket{f | V | F_0}}
  \right) V | \phi \ket
  \mbox{ .}
  \label{phi_tmp}
\end{equation}
Then we define $|\phi _1 \ket $ as
\begin{equation}
  | \phi_1 \ket \equiv  \frac{1}{1-G_1(E) V} | F_1 \ket
  \mbox{ ,}
  \label{phi_1}
\end{equation}
and (\ref{schr_ls}) can be written as 
\begin{equation}
  | \phi \ket  =  | \phi_1 \ket
    \frac{\braket{f | V | \phi }}{\braket{f | V | F_0}}
\mbox{ .}
    \label{phi_0}
\end{equation}
Multiplying (\ref{phi_0}) by $\bra f | V $ from the left,
we obtain the following condition to determine
the bound state energy $E$ and the wave function $|\phi \ket$:
\begin{equation}
  \Omega(E) \equiv
  \braket{f | V | F_0} - \braket{f | V | \phi_1}  =  0
  \mbox{ .}
  \label{phi_f}
\end{equation}
The bound state condition can be written
in terms of a continued fraction as 
\begin{equation}
  | \phi_i \ket \equiv \frac{1}{1 - G_i(E)V} | F_i \ket
\mbox{ ,}
  \label{phi_i}
\end{equation}
\begin{equation}
  | F_{i+1} \ket  =  G_{i}(E) V | F_{i} \ket
\mbox{ ,}
  \label{f_i} 
\end{equation}
and
\begin{equation}
   G_{i}(E)  =  G_{i-1}(E)
   - \frac{| F_{i} \ket \bra f | }{\braket{f | V | F_{i-1}}}
\mbox{ ,}
   \label{g_i}
\end{equation}
where $|\phi_i \ket$ is the $i$-th order wave functions.
The $i$-th Green's function $G_i(E)$
satisfies the orthogonality relation
\begin{equation}
  G_{i}(E) V | F_{j} \ket = 0
  \mbox{ .}
  \hspace{4mm} (i=j+1)
  \label{ortho}
\end{equation}
We rewrite (\ref{phi_i}) using (\ref{g_i}),
multiply $\bra f | V$ from the left, 
and obtain the following iteration formula for $| \phi_i \ket$:
\begin{equation}
  \braket{f | V | \phi_{i}} = \frac{\braket{f | V | F_{i}}^2}
  {\braket{f | V | F_{i}} - \braket{f | V | \phi_{i+1}}}
  \mbox{ .}
\label{MCFG_f_phi}
\end{equation}
As a result, the bound state condition (\ref{phi_f})
can be written in the form of a continued fraction:
\begin{equation}
  \Omega(E) =
  x_0 -
  \dfrac{x_{1}^2}
    {x_{1}-\dfrac{x_{2}^2}
          {x_{2}-\dfrac{x_{3}^2}
            {x_{3} {}-\,_{\ddots }}}} 
        \mbox{ .}
\end{equation}
Here, the matrix elements  $x_i$ are defined as
\begin{equation}
  x_i \equiv \braket{f|V|F_{i}}
  \mbox{ .}
  \label{mcf_x_i}
\end{equation}

In the MCF calculation, we only need to generate
$|F_i \ket$ and $G_i(E)$ iteratively,
following the iteration procedure (\ref{f_i}) and (\ref{g_i}).
In this procedure, $G_i(E)$ projects out the states
which are orthogonal to the lower order wave functions,
as indicated in (\ref{ortho}).
Therefore, $x_i$ is expected to become smaller for larger $i$,
and rapidly converges into zero.
The advantage of this continued fraction series based on
Ref. \cite{mcf_bind} for the Monte-Carlo calculation is
that the $x_i$ are simply given by the matrix element of
the potential between $|F_i \ket$ and the analytically
known function $|f \ket$, in contrast to those given in
Ref. \cite{mcf},
where we need the matrix elements between $|F_i \ket$. 

The bound state energy is determined by solving 
the equation
\begin{equation}
  \Omega(E)=0
\end{equation}
numerically.
In actual practice, we perform a parameter search for $E$.
Since we do not adopt any projections onto the ground state,
we can also apply MCF to study the excited states
without any restriction such as the $e^{-\tau H}$ projection.

Once the energy of the bound state has been determined,
the relevant eigenfunction is given as
\begin{equation}
  | \phi \ket =  
  {\cal N} \left[\,|F_1 \ket + y_1 | F_2 \ket + y_1 y_2 | F_3 \ket
    + y_1 y_2 y_3 | F_4 \ket + \cdots \right]
\mbox{ ,}
  \label{eigen_phi}
\end{equation}
where  $y_i $ is defined as
\begin{equation}
  y_i  \equiv    \frac{x_i}
  {x_i-\braket{f|V|\phi_{i+1}}} 
\mbox{ ,}
\end{equation}
and ${\cal N}$ is a normalization constant.

\section{Continued fractions and Monte-Carlo method}
The original form of the MCF iteration formula (\ref{f_i})
is not suitable for Monte-Carlo evaluations.
The difficulty is caused by the Green's functions $G_i(E)$.
Since the $G_i(E)$ obey the iteration formula
(\ref{g_i}) and are of a separable form,
\begin{equation}
   G_{i}(E)  =  G_{i-1}(E)
   - \frac{| F_{i} \ket \bra f | }{x_{i-1}}
\mbox{ ,}
\end{equation}
it does not seem practical
to generate the random numbers according to the
probability of $G_i(E)$.
In order to resolve these problems,
we use the orthogonality relation (\ref{ortho})
and modify the iteration formula (\ref{f_i}) as follows:
\begin{equation}
  |F_{i+1} \ket
  = G_0(E) V \left[ | F_{i} \ket - | F_{i-1} \ket
    \frac{x_{i}}{x_{i-1}} \right]
  \mbox{ .}
  \label{mod_f}
\end{equation}
The advantage of this formula is
that a higher order wave function $|F_{i+1} \ket $
is always generated by the free Green's function $G_0(E)$,
which has a simple analytical form.
Therefore, we can use the Monte-Carlo random walk technique
in the MCF iteration procedure.

However, it should be noted that
the iteration formula (\ref{mod_f}) is not of
the same form as the original Schr\"odinger equation.
The Schr\"odinger equation,
\begin{equation}
  |\phi \ket = G_0(E)V | \phi \ket
\mbox{ ,}
\label{diff_phi}
\end{equation}
can be regarded as a diffusion equation.
$|\phi \ket $ will converge to a ground state by iterating
(\ref{diff_phi}) with a Monte-Carlo random walk.
In contrast,  (\ref{mod_f}) can be written as 
\begin{equation}
  |\psi_1 \ket  = G_0(E) V | \psi_0 \ket
  \mbox{ ,}
  \label{mod_f_2}
\end{equation}
where $|\psi_1 \ket $ and $|\psi_0 \ket$
represent $|F_{i+1} \ket$ and
$[|F_{i} \ket - | F_{i-1} \ket  x_i / x_{i-1} ] $, respectively.
In our iteration formula (\ref{mod_f_2}),
the wave function $|\psi_1 \ket$ has 
different functional form from $|\psi_0 \ket$
and is generated from the two terms $|F_{i} \ket$
and $ | F_{i-1} \ket$.
Therefore, we cannot naively apply the random walk technique
of the diffusion equation.
Otherwise, a destructive interference of $|F_{i} \ket$
and $ | F_{i-1} \ket$ in  (\ref{mod_f})
may cause the ``sign problem''.

In this section, we  extend the random walk technique
for (\ref{mod_f_2}) and
propose a new method of counting weight.
The destructive interference
of two terms in (\ref{mod_f})
can be taken care of by our proper method of counting weight.
\subsection{Random walk in MCF}
We study an algorithm to apply the random walk method
with the free Green's function $G_0(R,R')$
to our following iteration formula of MCF:
\begin{equation}
  \psi_1(R) = \int G_0(R,R') W(R')\psi_0(R') dR'
  \mbox{ .}
\label{psi1_psi0}
\end{equation}
Here  we re-scale the variables so that
the free Green's function $G_0(R,R')$ is normalized as
\begin{equation}
  \left[
    -\nabla^{2}_{R} + 1 
  \right]
  G_0(R,R') = \delta(R-R')
\mbox{ ,}
\end{equation}
where $R$ denotes an $M$-dimensional
vector, i.e. $R=\{ z_1, \cdots , z_M \}$
defined in the standard manner given in Ref. \cite{kalos}.
The potential $W(R)$ is defined as
\begin{equation}
  W(R) \equiv -V(R)/B
\mbox{ ,}
\end{equation}
where $B$ is the binding energy. 

For simplicity, we assume that the wave functions
$\psi_1(R)$ and $\psi_0(R)$
and the potential $W(R)$ are positive definite.
Note that the positive sign of $W(R)$
implies that this is an attractive potential.
We assume $\psi_1(R)$ and $W(R)\psi_0(R)$ are represented by the sum of 
points $\{ R_{k,i} \}$ distributed with the probability $\psi_k(R)$
and the weights $\{ W_{k,i} \}$, where $k=0,1$.
The question is how $\{ R_{1,j} \}$ and $\{ W_{1,j}\}$
should be generated from given sets of $\{ R_{0,j} \}$ and $\{ W_{0,j}\}$.
The relation can be obtained by studying the following
overlap integral of  $\psi_1(R)$ with
an arbitrary function $F(R)$:
\begin{equation}
  {\cal I}  = \int F(R) \psi_1(R) dR
 \mbox{ .}
 \label{int_s1}
\end{equation}
${\cal I}$ is given by using $\{ R_{1,j} \}$ and $\{ W_{1,j}\}$ as
\begin{equation}
  {\cal I} = \sum_{j}^{N} W_{1,j} F(R_{1,j} )
 \mbox{ ,}
 \label{int_s2}
\end{equation}
while the same integral is expressed by $\{ R_{0,j} \}$ and $\{ W_{0,j}\}$
using (\ref{psi1_psi0}) as
\begin{equation}
 {\cal I}   = \int F(R) \sum_i^{N} W_{0,i} G_0(R,R_{0,i}) dR
  \mbox{ ,}
  \label{int_s3}
\end{equation}
where $N$ is the number of sampling points.

We multiply
the integrand of (\ref{int_s3})
by unity in the form $G_0(R,X)/G_0(R,X)$ and obtain
\begin{equation}
  {\cal I}   = \int A(R,X) G_0(R,X) dR 
  \mbox{ ,}
  \label{int_p2}
\end{equation}
where $A(R,X)$ is defined as
\begin{equation}
   A(R,X) \equiv  F(R) \sum_i^{N} W_{0,i} 
  \frac{G_0(R,R_{0,i})}{G_0(R,X)}
  \mbox{ .}
\end{equation}
Here, $X$ can be an arbitrary vector in $M$-dimensional space.
We impose hyper spherical boundary conditions on $G_0(R,X)$
for the bound state problem,
\begin{equation}
  G_0(R,X)=0 \hspace{3mm}\mbox{for}\hspace{3mm} |R|,|X| \rightarrow \infty
\mbox{.}
\end{equation}
$G_0(R,X)$ becomes a function of distance $|{\cal R}|$,
where ${\cal R}$ is defined as ${\cal R} \equiv R-X $.
Since $G_0(R,X)$ is normalized as
\begin{equation}
  \int G_0(R,X) dR =  \int G_0(|{\cal R}|) d{\cal R} = 1
  \mbox{ ,}
\label{g_0_1}
\end{equation}  
$G_0(|{\cal R}|)$ can be regarded as a probability function.

We change the integral variable from $R$ to ${\cal R}$ in (\ref{int_p2})
and generate a set of random numbers $\{ {\cal R}_j\}$ with probability
$G_0(|{\cal R}|)$.
The algorithm to generate $\{ {\cal R}_j\}$ in $M$ dimensional space is
given in Ref. \cite{kalos}.
Equation (\ref{int_p2}) is rewritten as 
follows:
\begin{equation}
  {\cal I}
  = \sum_j^{N} F(X+{\cal R}_j) 
  \frac{ \sum_i W_{0,i} G_0(X+{\cal R}_j,R_{0,i})}
  {N G_0(X+{\cal R}_j,X)}
  \mbox{ .}
  \label{int_p3}
\end{equation}
Since (\ref{int_p3}) holds for any values of $X$,
we choose $X$ as $R_{0,j}$ for the $j$-th random number ${\cal R}_j$.
Comparing (\ref{int_p3}) with (\ref{int_s2}),
we can define the new points $R_{1,j}$ and the weight $W_{1,j}$
as
\begin{equation}
  R_{1,j} \equiv  R_{0,j} + {\cal R}_j 
  \mbox{ ,}
  \label{MCFG_R1}
\end{equation}
and
\begin{equation}
  W_{1,j}  \equiv 
  \frac{ \sum_i W_{0,i} G_0(R_{0,j}+{\cal R}_j,R_{0,i})}
  {N G_0(R_{0,j}+{\cal R}_j,R_{0,j})}
\mbox{.}
  \label{MCFG_W1}
\end{equation}
As a result, (\ref{MCFG_R1}) and  (\ref{MCFG_W1})
show how the distribution of points $\{ R_{1,j} \}$ and
the weights $\{W_{1,j} \}$ of $\psi_1(R)$
should be generated from those of  $W(R) \psi_0(R)$.
A new point $R_{1,j}$ is generated by a random walk from $R_{0,j}$
similar to the diffusion equation.
However, a new weight 
$W_{1,i}$ is generated
by taking into account all of the old points $\{ R_{0,i} \}$
and weights $\{ W_{0,i} \}$ through the free Green's function.
In the naive random walk method, $W_{1,j}$ is simply given by $W_{0,i}$,
which projects into the ground state wave function. 
For a non-diffusion type integral formula like (\ref{psi1_psi0}),
weights of new points should be generated by using
this method of counting weight.

\subsection{Multiple cancellation}
In general, $\psi_0(R)$ and $W(R)$ are not positive definite,
and we must include their sign information in our formula.
This can be done by adding the sign $s_{0,j}$ to the weight $W_{0,j}$ as
\begin{equation}
  W_{0,j} \Rightarrow s_{0,j} W_{0,j} 
\mbox{ .}
\end{equation}
Similarly, adding the sign $s_{1,j}$ to  $W_{1,j}$,
(\ref{MCFG_W1}) is replaced by
\begin{equation}
  s_{1,j} W_{1,j} =
 \frac{ \sum_i s_{0,i}W_{0,i} G_0(R_{0,j}+{\cal R}_j,R_{0,i})}
 {N G_0(R_{0,j}+{\cal R}_j,R_{0,j})}
\mbox{ .}
\label{multi_cancel}
\end{equation}
Equation (\ref{multi_cancel}) is quite similar to 
the  ``multiple cancellation'' method proposed
by J. B. Anderson et al..\cite{cancel}
The multiple cancellation method with a random walk has been applied to
atomic physics, where it has provided very accurate results.
In the conventional random walk technique,
points with positive and negative signs walk independently.
Then the weight of each point grows,  
and a destructive interference of these wave functions
induces large numerical errors.
On the other hand, our formula (\ref{multi_cancel}) 
shows that weight and sign of new points are generated by
taking into account all the weights and signs of old points.
Therefore, the points with positive and negative weights
are canceled out in the course of counting the new weight.
Equation (\ref{multi_cancel}) plays an essential role
in the MCF calculation, and its importance will be demonstrated in
the one-dimensional example given in the next section.

\section{Applications to bound state problems}
The procedure of our method is summarized as follows:
\begin{enumerate}
\item Assume an appropriate form of the trial function $| F_0 \ket $.
\item Distribute the points by the Metropolis method\cite{metro}
  according to the functional form of  $| F_0 \ket $. 
\item Calculate the matrix element $x_i$ of the $i$-th iteration step.
\item Generate new points of $|F_{i+1} \ket$ by a random walk 
  according to the procedure described in the previous section.
\item Iterate steps 3 and 4 until the $x_i$ and $\Omega(E)$ converge.
\end{enumerate}

The binding energy $B(=-E)$ is determined by searching
for an energy that satisfies the condition $\Omega(E)=0$.

\subsection{One-dimensional model}
As a simplest example, we study a one-dimensional potential model.
In this one-dimensional problem, we can study the validity of our method by 
comparing with a conventional numerical method.
We take the mass of the particle as $m = 940$ MeV and
a simple central potential $V(x)$ of exponential
form with a one pion range as follows:
\begin{equation}
  V(x)  =  V_0 \exp(-a |x| )
\end{equation}
with
\begin{equation}
V_0 = -69.3 \hspace{1mm}(\mbox{MeV})
\mbox{ , }
a = 0.711 \hspace{1mm}(\mbox{fm}^{-1})\mbox{ .} 
\end{equation}
The strength of the potential is chosen so as to produce
three bound states at $-$38.7, $-$10.6 and $-$2.01 MeV.
The binding energy of the
second excited state is small, and is compared to
the deuteron binding energy. 
Since this model is one-dimensional and does not contain
any other degrees of freedom,
such as spin or isospin,
the wave function of the bound states can be
classified by their parity.
The ground and second excited states are even parity states,
and the first excited state is an odd parity state.
The validity of our method for the second excited state
is particularly interesting,
since it has the same symmetry as the ground state.

Introducing a dimensionless variable $z$ defined as
\begin{equation}
    z= \sqrt{2mB}x
\mbox{ ,}
\end{equation}
the free Green's function is given by 
\begin{equation}
  G_0(z,z')=\frac{1}{2}\exp(-|z-z'|)
\mbox{.}
\end{equation}
The trial function is taken to be a  Gaussian form as follows:
\begin{eqnarray}
  |F_0 \ket = | f \ket  & = & \frac{1}{\sqrt{b \sqrt{\pi}}}
  \exp(-\frac{x^2}{2b^2})  \hspace{1mm}
  \mbox{ for even parity,}\\
  & = & \sqrt{\frac{2}{b\sqrt{\pi}}}
  \frac{x}{b}\exp(-\frac{x^2}{2b^2})
  \hspace{1mm}\mbox{ for odd parity.}
\end{eqnarray}
We assume $b=3$ (fm) for the ground and first excited states
and $b=10$ (fm) for the second excited state.

In the Monte-Carlo calculation,
the numbers of sampling points were 20,000, 70,000 and 100,000
for the ground, first, and second excited states, respectively.
We needed a larger number of points 
for the excited states than for the ground state
in order to obtain reasonable accuracy.
The initial distribution of points with the probability 
$||F_0 \ket|$
was generated by the Metropolis method.
A typical number for the thermalization step was $100$ in this problem.   

\begin{figure}[htbp]
    \epsfxsize = 15 cm
\centerline{\epsfbox{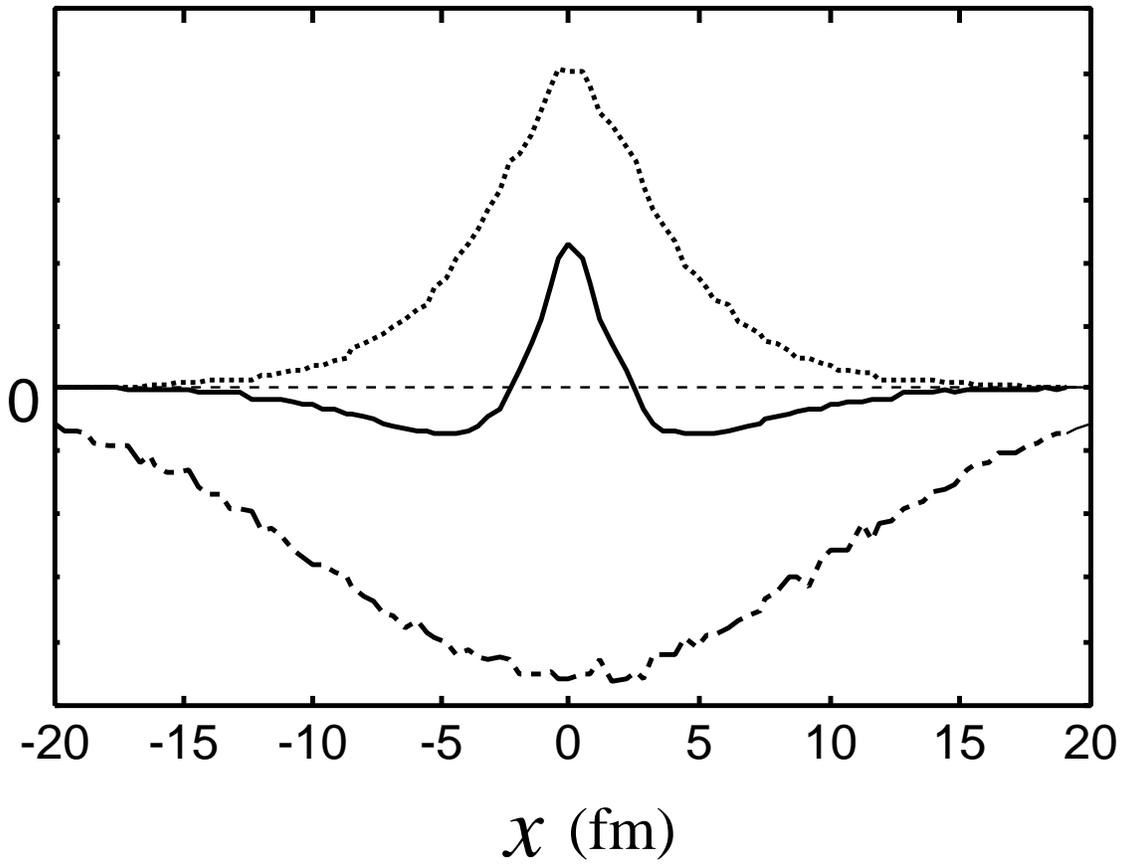}}
\caption{MCF wave functions
  at  energy (2.01 MeV) of the 2nd-excited state.
  The solid line, dotted line, and dashed line correspond to
  $|F_2 \ket$, $|F_1 \ket$, and $-|F_0 \ket x_1 / x_0$,
  respectively.}
\label{fig:2}
\end{figure}

\begin{figure}[htbp]
\epsfxsize = 15 cm
\centerline{\epsfbox{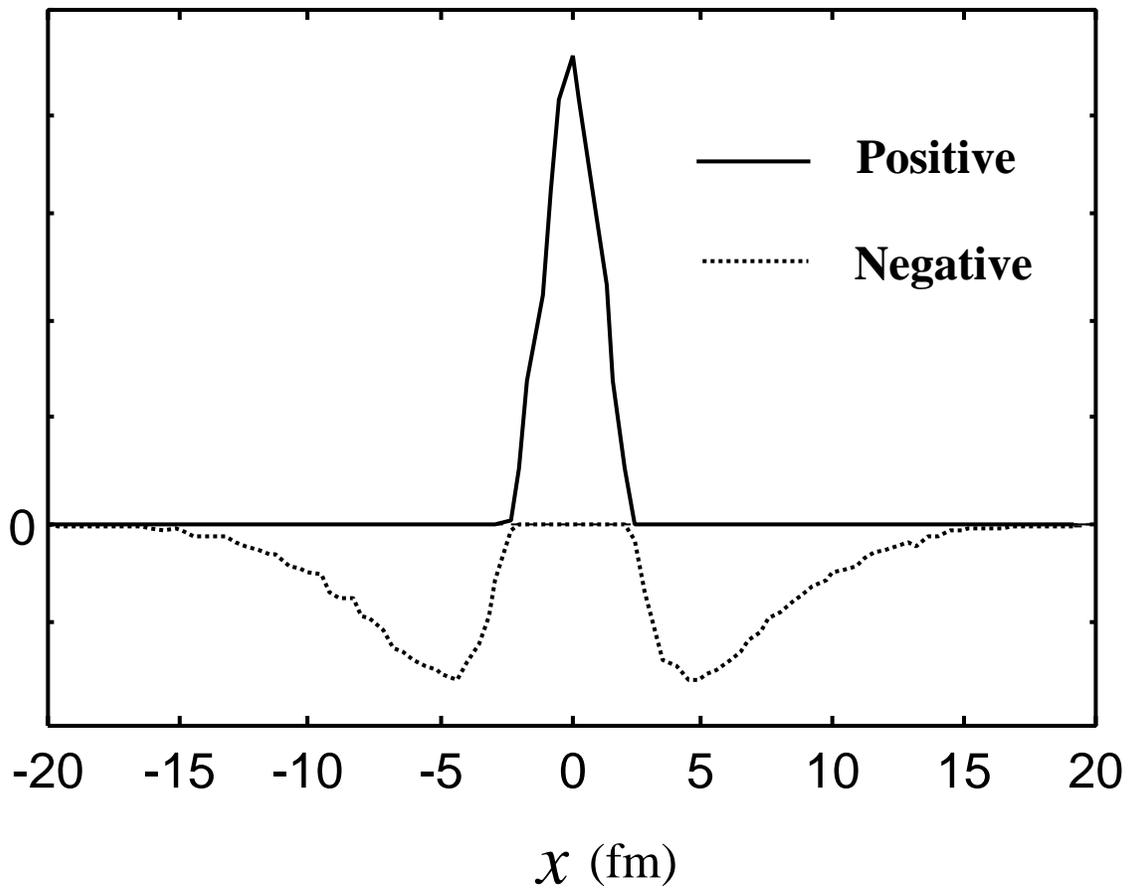}}
\caption{MCF wave function $|F_2 \ket$
  generated by our method of weight counting.
  The solid and dotted lines correspond to the population of
  the points for  positive and
  negative weight, respectively.}
\label{fig:6}
\end{figure}

The fast convergence of the MCF iteration is related to the
orthogonality relation in (\ref{ortho}),
which is implemented in our iteration formula in (\ref{mod_f}).
As an example,
we consider the second order MCF wave function $|F_2 \ket $,
\begin{equation}
  |F_2 \ket = G_0(E)V
  \left[
    |F_1 \ket - |F_0 \ket\frac{x_1}{x_0}
  \right]
  \mbox{ .}
\end{equation}
$|F_2 \ket $ is
generated by  two terms,
$|F_1 \ket $ and $-|F_0 \ket x_1 / x_0 $,
which will cancel each other.
The higher order $|F_i \ket $ decreases quickly
with this cancellation mechanism.
Therefore, accurate treatment of this cancellation is important.
If we generate $|F_2 \ket$ from two terms independently,
the resultant distribution of the points for $|F_2 \ket$ 
is given by the large overlap of the points
with positive and negative signs shown in Fig. {\ref{fig:2}}
at the energy of the second excited state.
$|F_2 \ket$ generated from these points will include
large errors, which is just our ``sign problem''.
Because of this sign problem, 
it is difficult to obtain
convergence for the higher order MCF wave functions in the
standard random walk method.
On the other hand,
using our method of counting weight in (\ref{multi_cancel}),
the cancellation between two terms is performed analytically,
and the distribution of the positive points for 
$|F_2 \ket $ is clearly separated from the negative points,
as shown in Fig. \ref{fig:6}.
Our method of counting weight plays an important role
in MCF iteration with the Monte-Carlo method.

\begin{figure}[htbp]
\epsfxsize = 15 cm
\centerline{\epsfbox{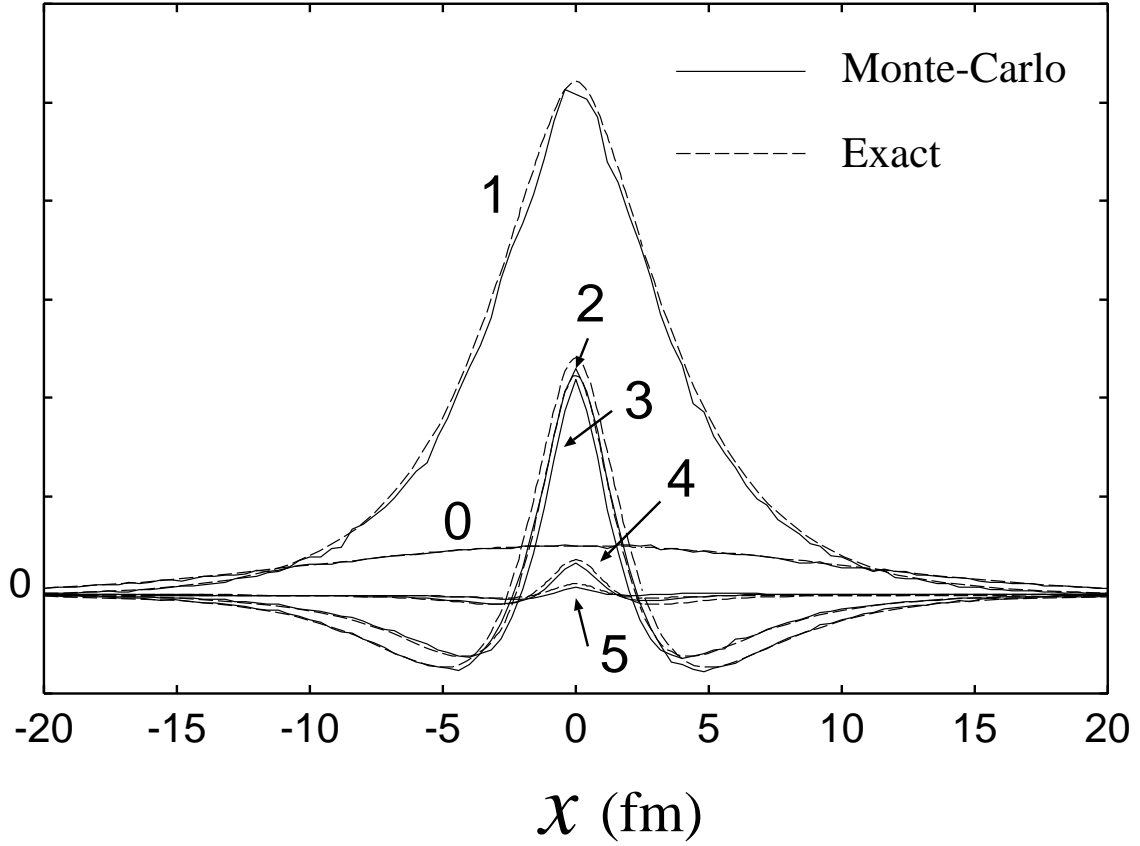}}
    \caption{The wave function of MCF, $|F_i \ket$,
at the energy of the 2nd-excited state ($-2.01$ MeV).
The solid lines are obtained by Monte-Carlo method, and
the dashed lines are obtained by a numerically `exact' method.
The numbers $(0 \sim 5)$
specify the wave functions $|F_i \ket$.
Normalization of the wave functions obtained by the Monte-Carlo method
is adjusted to the `exact' ones.}
\label{fig:5}
\end{figure}

MCF wave functions at each iteration step are shown in Fig. \ref{fig:5}.
They decrease rapidly as the iteration steps increase,
even for the second excited state.
This agrees well with the results obtained using
the standard method of numerical integration.

\begin{figure}[htbp]
  \epsfxsize = 15 cm
  \centerline{\epsfbox{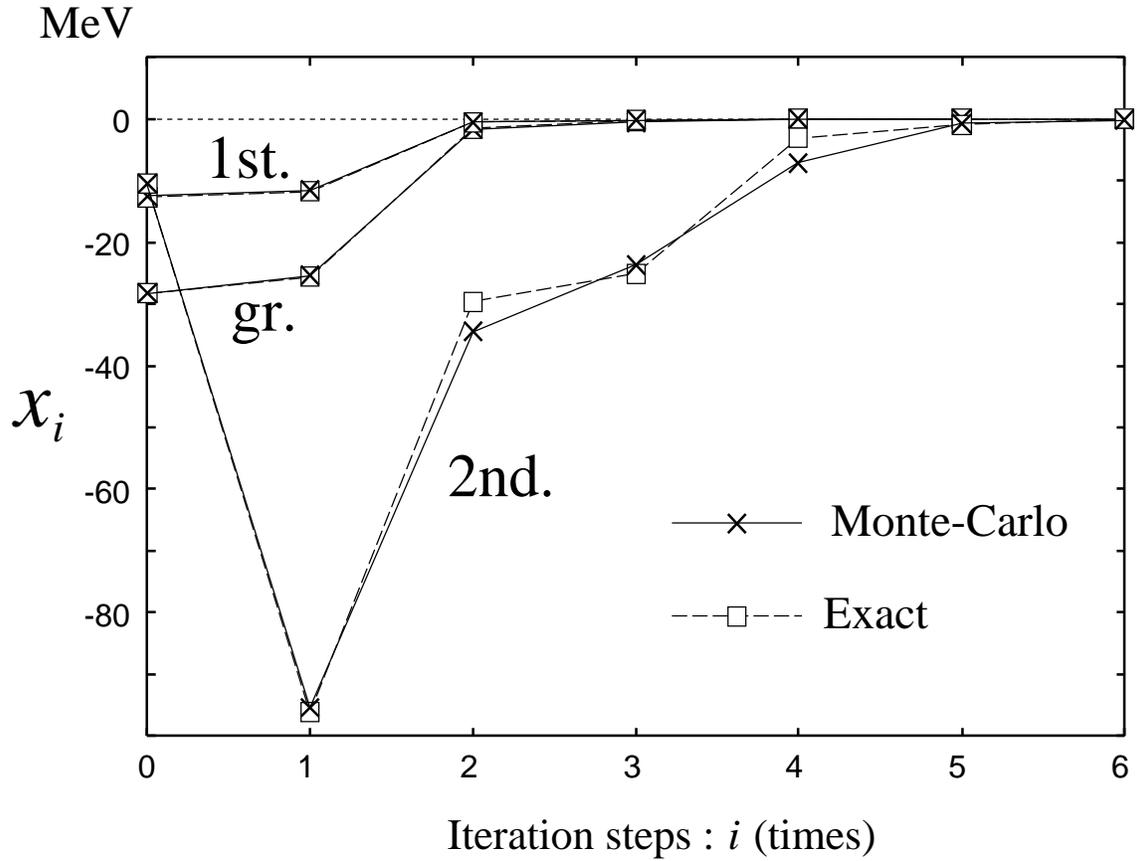}}
  \caption{Convergence of $x_i$ as a function of the iteration step $i$
    for the ground, 1st-, and 2nd-excited states in units of MeV.
    Cross points were obtained by the Monte-Carlo method,
while the box points were obtained by a numerically 'exact' method.}
\label{fig:1}
\end{figure}
\begin{figure}[htbp]
  \epsfxsize = 15 cm
\centerline{\epsfbox{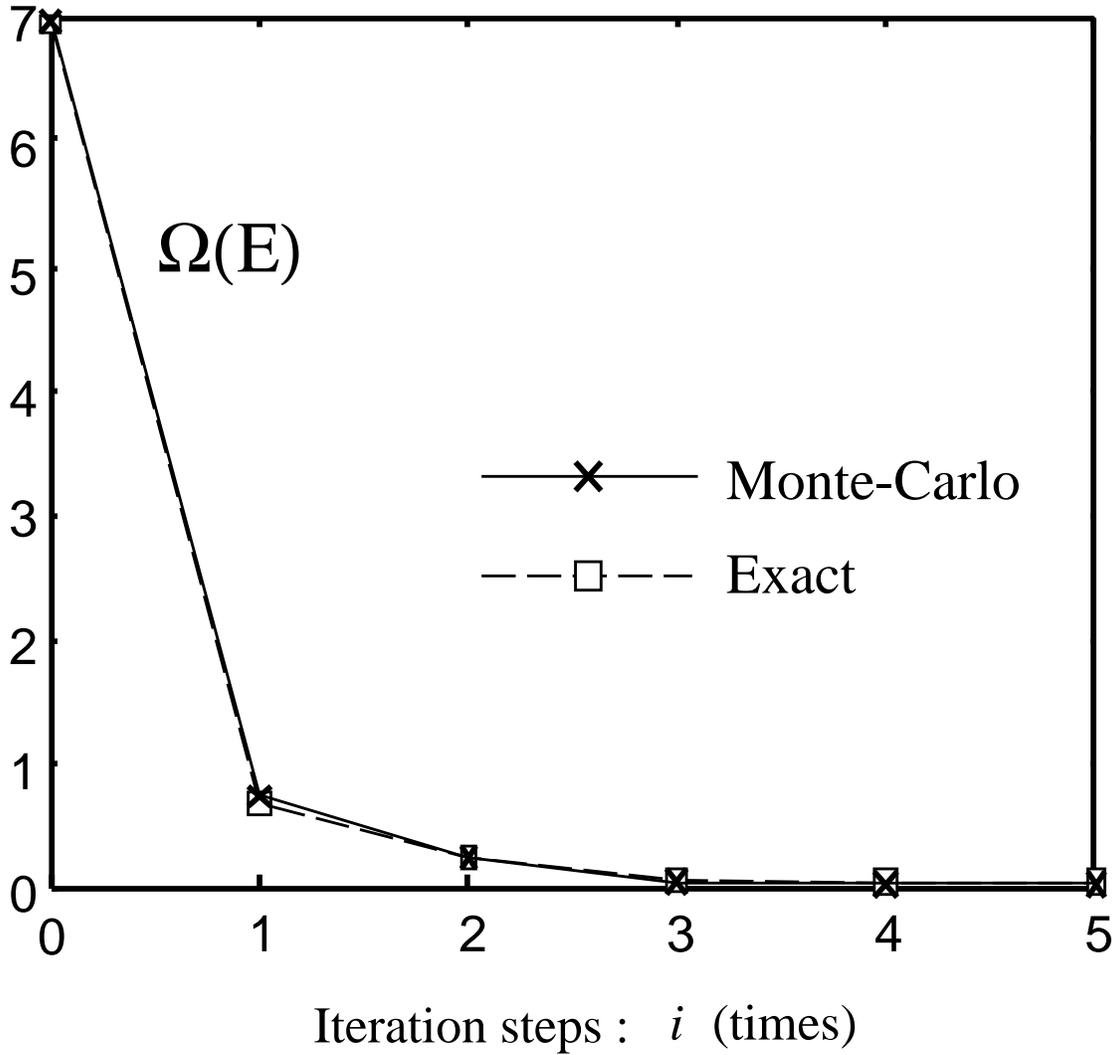}}
    \caption{Convergence of $\Omega(E)$ as a function of the iteration
step $i$ at the ground state energy with arbitrary scale.
The points marked by crosses
and boxes were obtained by the Monte-Carlo method
and a numerically `exact' methods, respectively.}
\label{fig:3}
\end{figure}

Consequently, the matrix elements $x_i$ defined in (\ref{mcf_x_i})
and $\Omega(E)$ also converge rapidly
as a function of the iteration step  as shown 
in Fig. \ref{fig:1} and \ref{fig:3}, respectively.
A few steps of iteration is enough to obtain reasonable convergence
for the ground and first excited states.
Slower convergence for the second excited state is due to
our simple Gaussian trial function,
even though one expects nodes in the wave function of the second excited state.

\begin{figure}[htbp]
  \epsfxsize = 15 cm
\centerline{\epsfbox{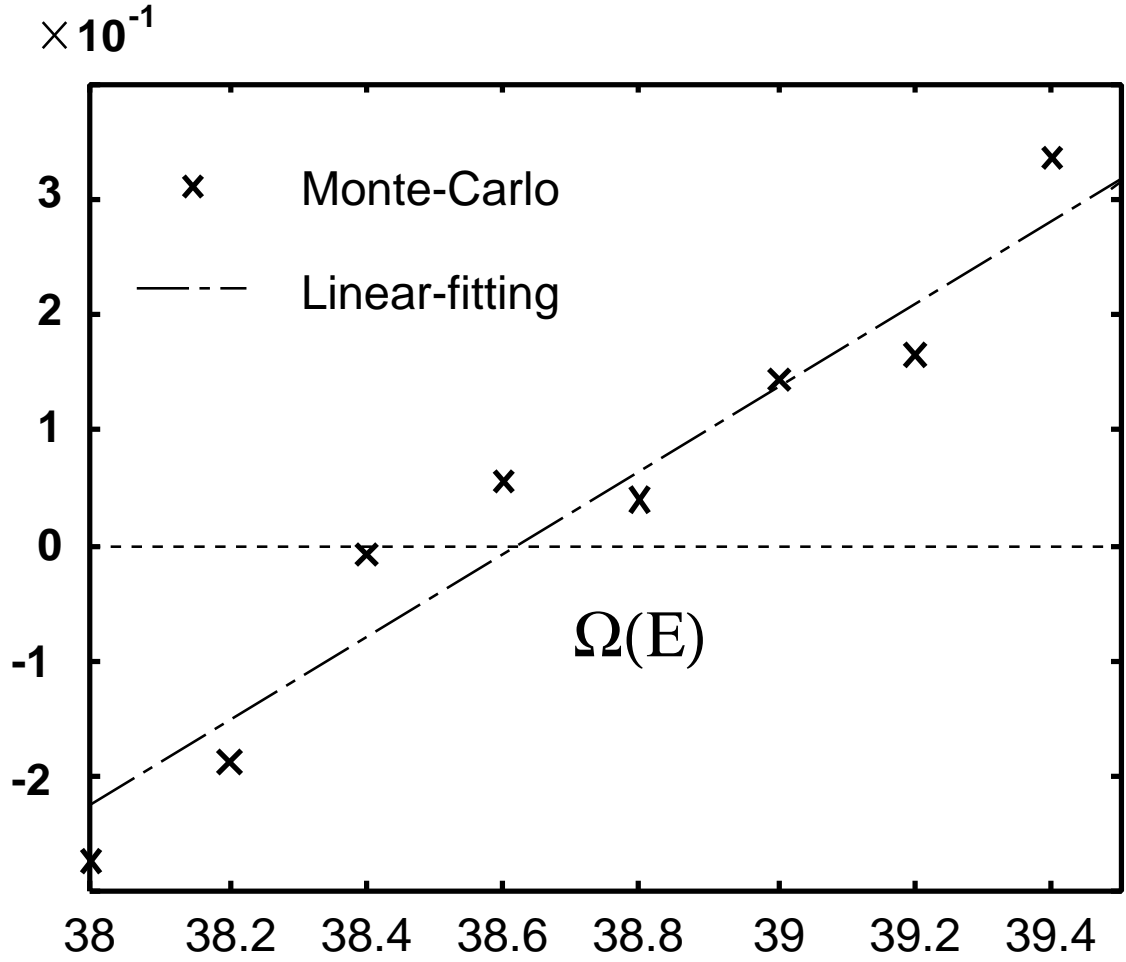}}
\caption{Energy dependence of $\Omega(E)$.
Points marked by crosses were obtained by the Monte-Carlo method.
The dotted-dashed line was obtained by 
  a least square fitting.
The units of the vertical axis are the same as in  Fig. 5.}
    \label{fig:4}
\end{figure}

The energy dependence of  $\Omega(E)$ is shown in Fig. \ref{fig:4}
for the ground state.
To determine the binding energy from the condition $\Omega(E)=0$, 
we fitted $\Omega(E)$ by using
the linear energy-dependence of $\Omega(E)=0$
near the binding energy with the least square search method.
Since values of $\Omega(E)$ fluctuate due to sampling errors,
we performed calculations several times with different sets of random numbers.
Averaging these results,
we obtained the energies of three bound states and errors estimated by the 
standard deviation, as shown in Table \ref{table:1}.
Our results with the Monte-Carlo method are in a good agreement with those of
the `exact' calculation for all three states.

\begin{figure}[htbp]
\epsfxsize =  15 cm
\centerline{\epsfbox{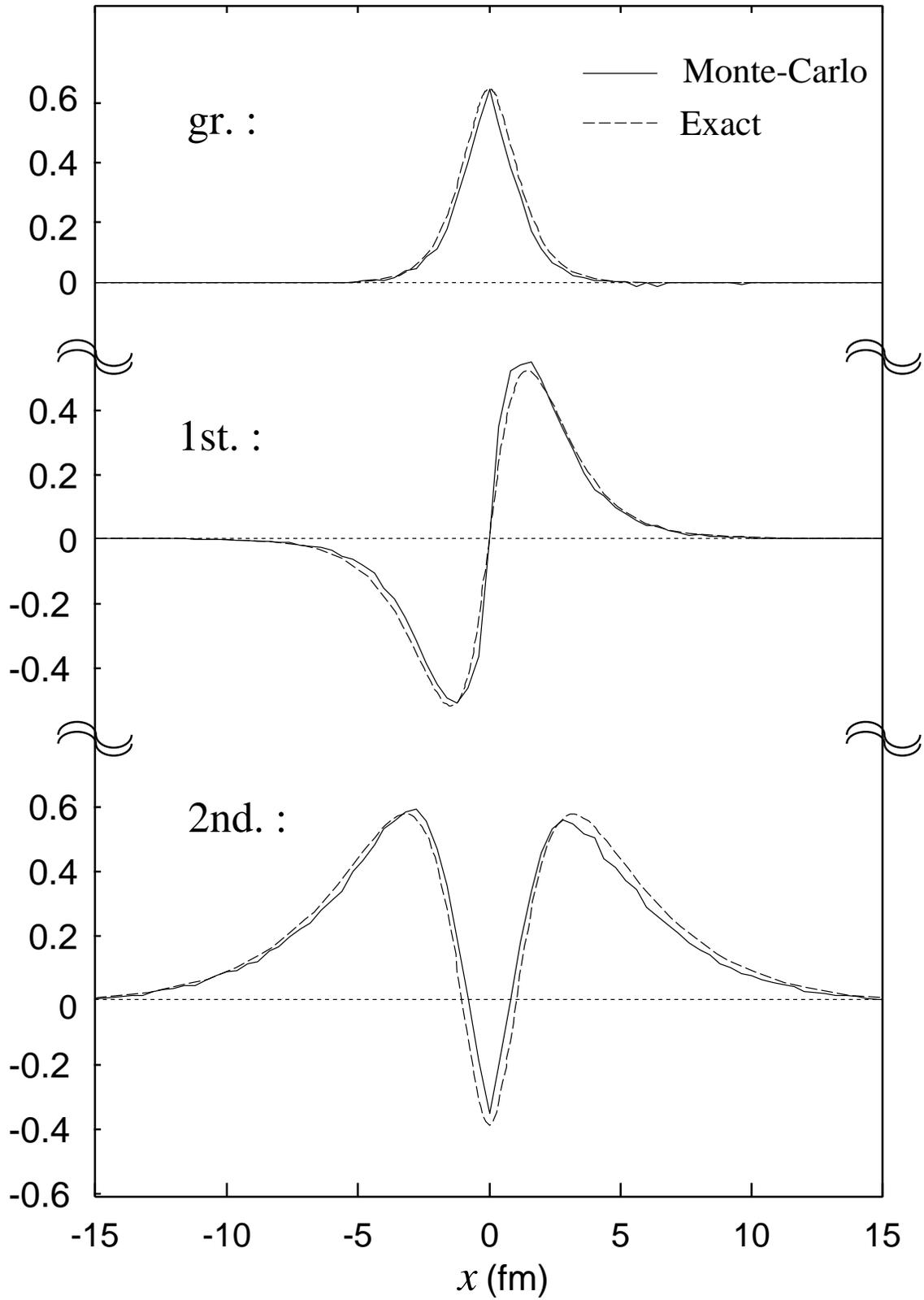}}
\caption{The eigenfunction
  for the ground, 1st-, and 2nd-excited states.
The solid lines were obtained by the Monte-Carlo method.
The dashed lines were obtained by a numerically `exact' method.
Normalization of the wave functions obtained by the Monte-Carlo method
were adjusted to the exact ones.}
\label{fig:7}
\end{figure}

Once we obtained the binding energies,
the eigenfunctions were given by the MCF wave functions $|F_i \ket $
and matrix elements $ x_i$.
The wave functions for the ground, first, and second excited
states are shown in Fig. \ref{fig:7}.
The wave functions are also in good agreement with the `exact' calculation.
Note that we started with a trial function
without nodes for the second excited state.
We could use more realistic trial functions with nodes,
and if we did so we would obtain faster convergence.
However, our result shows that
it is not necessary to know the correct nodes of the wave function 
when we construct a trial function in this method.
In summary, our new Monte-Carlo method works
for the calculation of binding energies and wave functions
including the excited states,
and we believe that our method
will be  useful  in studying few-nucleon bound state systems.
\subsection{Four-nucleon bound state as an example of
           a multi-dimensional problem}
As a realistic example,
we study a four-nucleon bound state in three-dimensional space.
The Schr\"{o}dinger equation is expressed as follows:
\begin{eqnarray}
  -\sum_{k=1}^{3} \frac{1}{2\mu_k}
  \nabla_{r_k}^2 | \phi \ket
  + \sum_{i<j}^{4} V_{ij} | \phi \ket = -B | \phi \ket
\mbox{ .}
  \label{9-dim_sch}
\end{eqnarray}
Here, the \bv{r}$_k$ are the Jacobi coordinates defined as
\[
\left\{
\begin{array}{lcl}
  \bv{r}_1 & = &
  \bv{x}_2 - \bv{x}_1 \mbox{ ,}\\
  \bv{r}_2 & = &  \bv{x}_3
  - (\bv{x}_1 + \bv{x}_2)/2 \mbox{ ,}\\
  \bv{r}_3 & = &  \bv{x}_4
  - (\bv{x}_1+ \bv{x}_2+ \bv{x}_3)/3
  \mbox{ ,}
\end{array}
\right. \]
and the $\mu_k$ are reduced masses
corresponding to the coordinates $\bv{r}_k$:
\[\left\{
\begin{array}{lcl}
\mu_1 & = & m/2 \mbox{ ,}\\
\mu_2& = & 2m/3 \mbox{ ,}\\
\mu_3 & = & 3m/4 \mbox{ .}
\end{array}
\right. \]

Here, we adopt the Volkov potential, which is a central potential
and has a soft core at short range:\cite{volkov}
\begin{equation}
  V_{ij}  =  V_1 \exp(-a_1 r_{ij}^2)
  + V_2 \exp(-a_2 r_{ij}^2)
\mbox{ ,}
\end{equation}
with 
\[\left\{
  \begin{array}{ll}
    V_1 = -83.34 \mbox{ (MeV) ,} &
    a_1 = 1.60^{-2} \mbox{ (fm$^{-2}$) ,}\\ 
    V_2 = 144.86 \mbox{ (MeV) ,} & 
    a_2 = 0.82^{-2} \mbox{ (fm$^{-2}$) ,}
    \mbox{ .}
  \end{array}
\right. \]
Since the ground state energy of $^4$He using this potential 
has been studied by various methods
such as the Faddeev-Yakubovsky (FY) method 
taking into account only the $S$-wave component,\cite{FY}
the stochastic variational method (SVM)\cite{SVM} and
the hyperspherical harmonics (HH),\cite{HH}
it is a good example to examine our method in a four-body system.

The free Green's function in dimensionless coordinates is expressed as
\begin{equation}
  G_0(R,R') = \frac{1}{2(2\pi)^{4}}
  \frac{1}{{\cal R}^{4}}
  \left(
    1 + \frac{6}{{\cal R}}
    +\frac{15}{{\cal R}^2} + \frac{15}{{\cal R}^3}
  \right)
  \exp(-{\cal R})
  \mbox{ ,}
\end{equation}
where ${\cal R} \equiv |R-R'|$ is a relative distance
in the nine-dimensional space.
As the dimensionality increases, the
free Green's function becomes more singular 
at ${\cal R}=0$.
It is rather difficult to treat the singularity of 
the free Green's function with the conventional integration method.
Since the random walk method inverts the free Green's function
analytically, we can obtain numerically stable results.
A standard method to generate points with the probability 
$G_0({\cal R})$ is given in Ref. \cite{kalos}.

The ground state of $^4$He is expected to be spatially symmetric
and anti-symmetric in spin-isospin space.
Then we take a trial function $|F_0 \ket$
given by the product of a Gaussian wave function
$| \varphi_s \ket$, a correlation function $g(r)$
and the spin-isospin function $| \Xi \ket$, as
\begin{equation}
  | F_0 \ket  = {\cal N} \prod_{i<j} g(r_{ij})
  | \varphi_s \ket |\Xi \ket
  \mbox{ ,}\\
\end{equation}
with 
\begin{equation}
 | \varphi_s \ket = \prod_{i<j} \exp(-r_{ij}^2 / 2b_0^2) 
  \mbox{ ,}
  \label{try_Gauss}
\end{equation}
where $b_0$ is a size parameter.
We used the form of correlation function $g(r)$
\begin{equation}
  g(r_{ij})  =  [ 1 - \alpha \exp(-r_{ij}^2/b_g)]^2
  \label{try_corr}
\end{equation}
with 
\[
\alpha = 0.11 \mbox{ , } b_g = 0.5 \mbox{ (fm$^{2}$)}
\mbox{ .}
\]
The parameters of the correlation function are determined to
simulate the wave function of a two-nucleon bound state
in this potential.
The size parameter $b_0$ of the Gaussian wave function in (\ref{try_Gauss})
is simply determined by the variational method with one parameter.
For the trial function $| f \ket $,
we assume the same functional form as $| F_i \ket$,
and take a slightly larger value of the size parameter $b_0$,
\begin{eqnarray}
  b_0 & = & 3.0 \mbox{ (fm) : for } | F_0 \ket \\ 
    & = & 4.5 \mbox{ (fm) : for }  | f \ket  \mbox{ .}
\end{eqnarray}
Since the Volkov potential is a central potential with no exchange terms,
the symmetry of the spatial and spin-isospin
wave functions are preserved separately.

\begin{figure}[htbp]
\epsfxsize = 15 cm
\centerline{\epsfbox{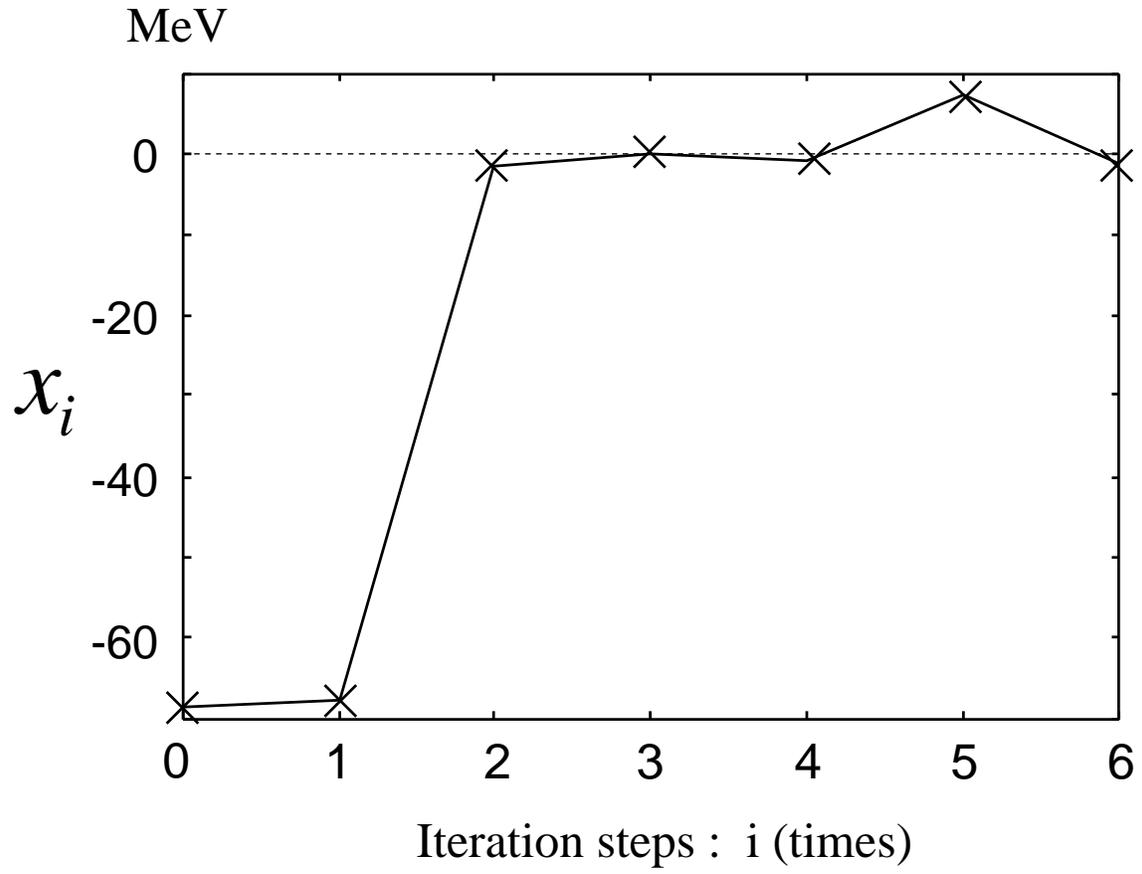}}
    \caption{Convergence of $x_i$ as a function of the
iteration step $i$ at $B=30$ (MeV).}
\label{fig:10}
\end{figure}

\begin{figure}[htbp]
  \epsfxsize = 15 cm
\centerline{\epsfbox{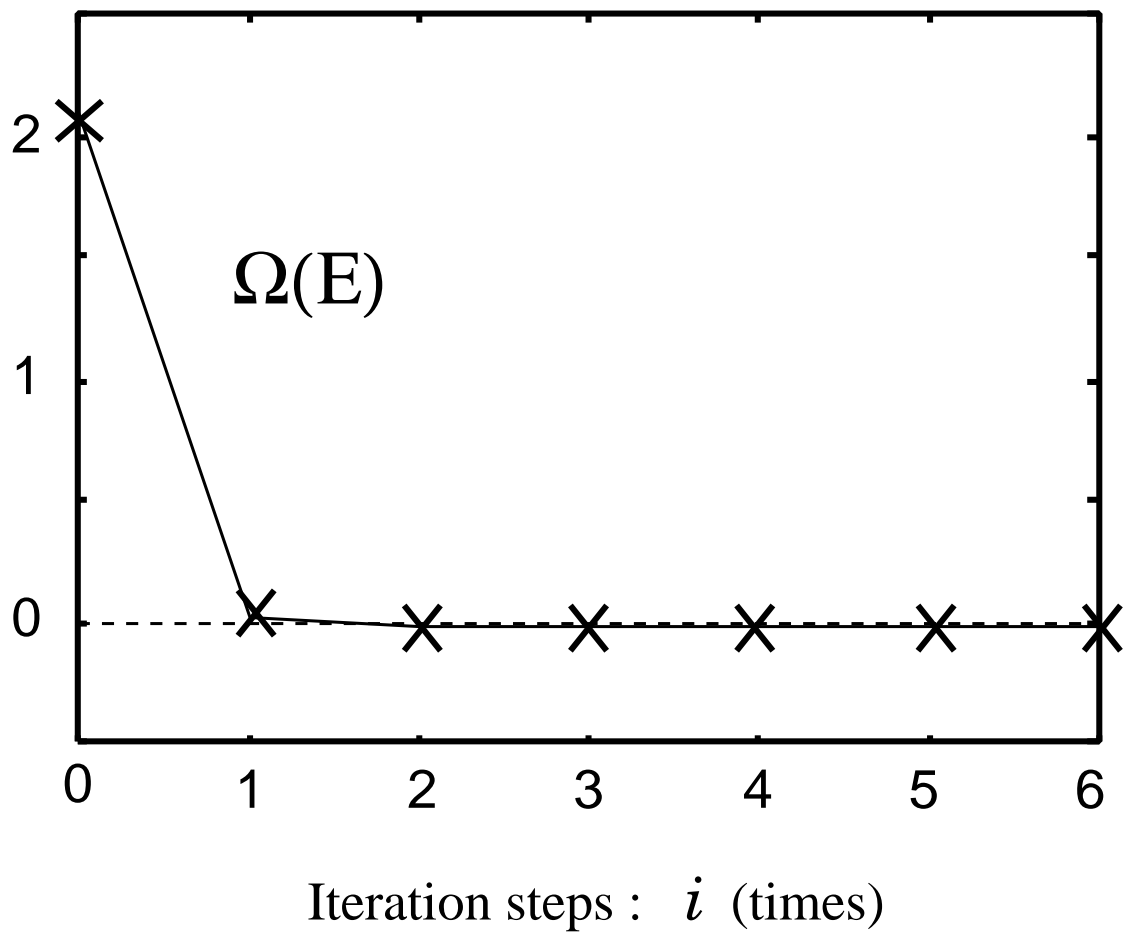}}
    \caption{Convergence of $\Omega(E)$ as a function of the
      iteration step $i$ at $B=$30 (MeV).}
\label{fig:11}
\end{figure}

The convergence of $x_i$ is displayed in Fig. \ref{fig:10},
which shows that three-times
iterations are enough in this example.
Similarly, $\Omega(E)$ also converges
rapidly, as shown in Fig. \ref{fig:11}.
Here, we take $N=200,000$ sampling points.

\begin{figure}[htbp]
\epsfxsize = 15 cm
\centerline{\epsfbox{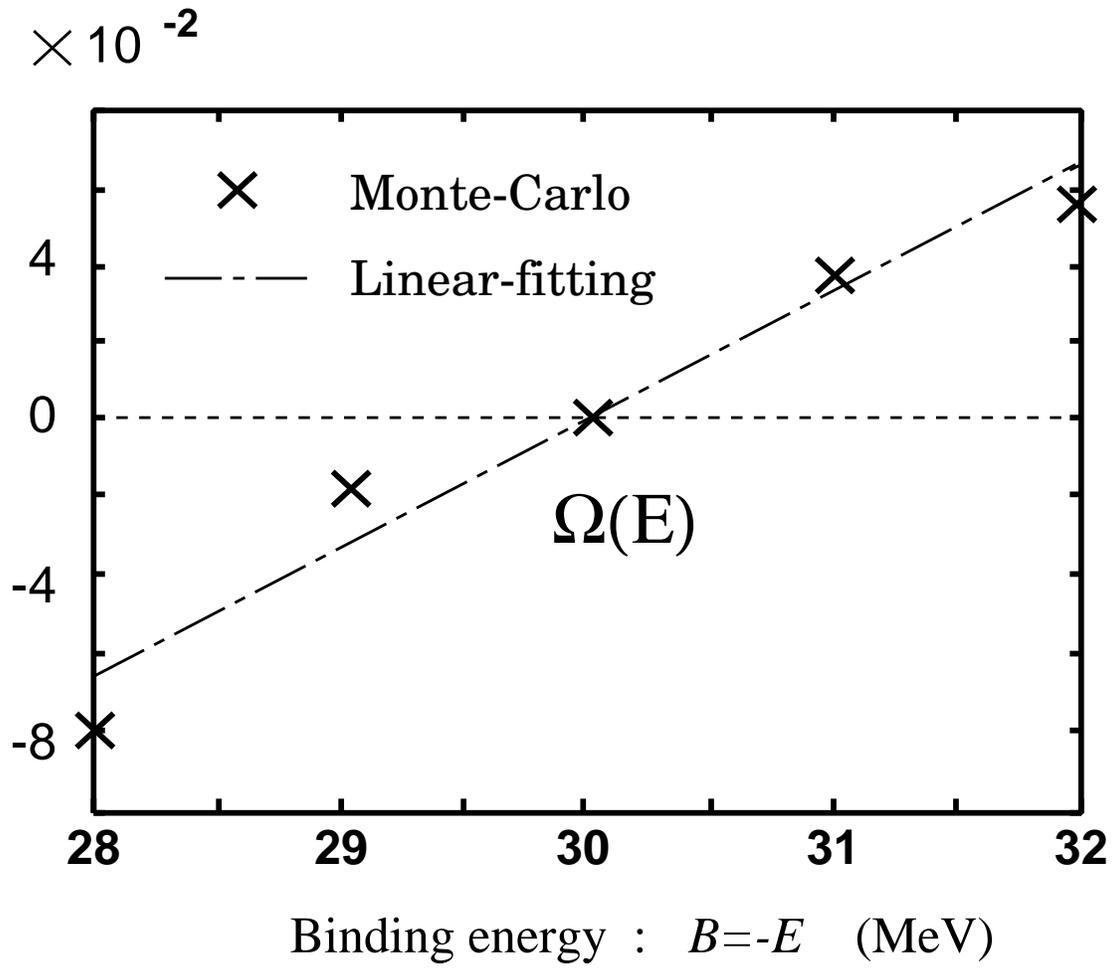}}
\caption{Energy dependence of $\Omega(E)$,
  fitted using the same method as in the one-dimensional model.}
\label{fig:12}
\end{figure}

The bound state energy and its uncertainty are estimated
in the same way as in the one-dimensional case.
We assume a linear function for fitting
the energy dependence of $\Omega(E)$,
as shown in Fig. \ref{fig:12}.
As a result, we obtain $E=-29.88 \pm 0.59$ (MeV).
Our results for the binding energy of $^4$He are in good agreement
with the results of the other methods within uncertainties,
as tabulated in Table \ref{table:2}.
It is shown that our method also works for higher 
dimensional examples.

\section{Summary and discussion}
The Monte-Carlo method is useful in the study of 
the bound state of the many-nucleon systems
because of its simplicity and efficiency
for multi-dimensional calculations.
In this work, we propose a new method to study
the bound state of a few-nucleon system without using the
diffusive-type projection method.
Therefore it is expected to work not only for the ground state
but also for excited states.

We used the method of continued fractions (MCF)
as the main formalism of the calculation.
The advantages of this method are its ability to calculate
the excited states
and the rapid convergence of the continued fraction series.
We modified the original formula of MCF into a suitable form
for the Monte-Carlo random walk method.
We generated MCF wave functions by the random walk method.
Here, it was shown that the cancellation mechanism of
lower order MCF wave functions is important,
and we should use our method  of counting weight in this calculation
to avoid the ``sign problem''. 

Our method was examined in application to
a one-dimensional problem in detail.
We demonstrated that our new method works even for 
the calculation of excited states.
We also studied the bound state of a four-nucleon system
and obtained results that compare well with those of other methods.
These facts suggest that 
our method is a powerful method to study the bound states of 
few-nucleon systems.

Application of this method to few-nucleon systems
with realistic nuclear potentials including exchange terms
is one of our future problems.
Here, we have to deal with the coupled channel problem
of wave functions with  various types of exchange symmetry.
It is important to retain the correct symmetry of the wave function
in each iteration step.
It is also necessary to take into account the correlation of nucleons 
in the trial function as much as possible. 
The wave function obtained by a fully variational calculation 
should be used as a more plausible trial function in our procedure.
This will also  accelerate the convergence of the continued fraction series.

\section*{Acknowledgements}

The authors would like to thank Professor H.~Ohtsubo for
many critical discussions and comments.

\newpage
\newpage
\newpage
\newpage
\newpage
\newpage
\newpage
\newpage
\newpage
\newpage
\begin{table}[htbp]
\bct
    \leavevmode
\begin{tabular}{lrr}
\hline
\hline
state & `exact'  & Monte-Carlo \\
\hline
ground & $-38.7$  & $-38.8$ $\pm$ 0.1 \\
\hline
1st-excited & $-10.6$  & $-10.6$ $\pm$ 0.1 \\
\hline
2nd-excited &  $-2.01$ & $-2.03$ $\pm$ 0.14 \\
\hline
\end {tabular}
\ect
\vspace{3mm}
\caption{The eigenvalues (MeV)
  for the ground, 1st-, and 2nd-excited states.
  The results obtained by the Monte-Carlo method 
  are compared with those obtained by a numerically `exact' method.}
\label{table:1}
\end{table}
\newpage
\begin{table}[htbp]
\bct
    \leavevmode
\begin{tabular}{ll}
\hline
\hline
Method & Energy (MeV) \\
\hline
Faddeev-Yakubovsky \cite{FY}& $-30.27$   \\
\hline
HH \cite{HH}& $-30.40$  \\
\hline
SVM \cite{SVM}&  $-30.42$ \\
\hline
MCF with MC & $-29.88$ $\pm$ 0.59\\
\hline
\end {tabular}
\ect
\vspace{3mm}
\caption{The bound state energy of $^4$He.}
\label{table:2}
\end{table}

\end{document}